\documentclass[twocolumn]{aa} 
\usepackage{graphicx}
\usepackage{txfonts}
\newcommand{\Msuno}{\hbox{$\hbox{M}_\odot$}}

\newcommand{\kms}{\hbox{${\rm km}\:{\rm s}^{-1}\;$}}
\newcommand{\kmso}{\hbox{${\rm km}\:{\rm s}^{-1}$}}
\newcommand{\teff}{$T_{\rm eff}\;$}  
\newcommand{\teffo}{$T_{\rm eff}$}  
\newcommand{\logg}{$\log{g}\;$}   	
\newcommand{\loggo}{$\log{g}$}  	

\begin{document}
\title{Chemical abundances of late-type pre-main sequence stars\\ in the
$\sigma$~Orionis cluster}
\titlerunning{Metallicity of the $\sigma$~Orionis cluster}
\authorrunning{Gonz\'alez Hern\'andez et al.}
%
%
\author{J.~I.~Gonz\'alez Hern\'andez\inst{1,2},  
J.~A.~Caballero\inst{3,4}, R.~Rebolo\inst{4,5},
V.~J.~S.~B\'ejar\inst{4}, D.~Barrado~y~Navascu\'es\inst{6},
E.~L.~Mart\'{\i}n\inst{4}, \and M.~R.~Zapatero~Osorio\inst{4}
}
\offprints{J.~I. Gonz\'alez Hern\'andez.}
\institute{
Observatoire de Paris-Meudon, GEPI, 5 place Jules
Janssen, 92195 Meudon Cedex, France \\
\email{Jonay.Gonzalez-Hernandez@obspm.fr}
\and
Cosmological Impact of the First STars (CIFIST) Marie Curie Excellence Team
\and
Dpto. de Astrof\'{\i}sica y Ciencias de la Atm\'osfera, Facultad de F\'{\i}sica,
Universidad Complutense de Madrid, E-28040 Madrid, Spain 
\and
Instituto de Astrof{\'\i }sica de Canarias, E-38205 La Laguna,
Tenerife, Spain 
\and
Consejo Superior de Investigaciones Cient{\'\i }ficas, Spain
\and
Laboratorio de Astrof\'{\i}sica Espacial y F\'{\i}sica Fundamental,
INTA, P.O. Box 78, E-28691 Villanueva de la Ca\~nada, Madrid,
Spain  
} 
%
\date{Received .. June 2008; accepted .. July 2008}

 
\abstract
{The young $\sigma$~Orionis cluster is an important location for
understanding the formation and evolution of stars, brown dwarfs, 
and planetary-mass objects.
Its metallicity, although being a fundamental parameter, has not been well
determined yet.} 
{We present the first determination of the metallicity of nine
young late-type stars in $\sigma$~Orionis.}
{Using the optical and near-infrared broadband photometry available in
the literature we derive the effective temperatures for these nine
cluster stars, which lie in the interval 4300--6500 K (1--3 \Msuno).
These parameters are employed to compute a grid of synthetic spectra
based on the code MOOG and Kurucz model atmospheres. We employ a
$\chi^2$-minimization procedure to derive the 
stellar surface gravity and atmospheric 
abundances of Al, Ca, Si, Fe, Ni and Li, using multi-object optical
spectroscopy taken with WYFFOS+AF2 at at the William Herschel Telescope (
$\lambda/\delta\lambda\sim7500$).}  
{The average metallicity of the $\sigma$~Orionis cluster is
[Fe/H] $ = -0.02\pm0.09\pm0.13$ (random and systematic errors).
The abundances of the other elements, except lithium, seem to be 
consistent with solar values. Lithium abundances are in agreement with the
``cosmic'' $^7$Li abundance, except for two stars which show a $\log
\epsilon(\mathrm{Li})$ in the range 3.6--3.7 (although almost consistent within
the error bars). There are also other two stars with $\log
\epsilon(\mathrm{Li})\sim 2.75$. We derived an average radial velocity
of the $\sigma$~Orionis cluster of $28\pm4$\,\kmso.} 
{The $\sigma$~Orionis metallicity is roughly solar.} 


\maketitle

\section{Introduction}
\label{introduction}

The \object{$\sigma$~Orionis cluster} is a nearby, very young region that has
become a suitable place for searching for and characterizing
substellar objects. 
The age has been estimated at $3 \pm 2$\,Ma (Oliveira et~al. 2002;
Zapatero Osorio et~al. 2002a; Sherry et~al. 2004) and its heliocentric
distance is roughly $360^{+70}_{-60}$\,pc (Brown et~al. 1994). 
The cluster is relatively free of extinction ($A_V \ll$ 1\,mag --
Lee 1968; B\'ejar et~al. 2004) and has a moderate spatial member
density (Caballero 2008a) and {\bf a large frequency of
intermediate-mass stars with discs}.  
A compilation of different determinations of the age, distance, and disc
frequency at different mass intervals is provided in Caballero (2007). 
The cluster contains several dozen brown dwarfs with spectroscopic features of
youth and with discs (B\'ejar et~al. 1999; Zapatero Osorio et~al. 2002a;
Barrado~y~Navascu\'es et~al. 2002; Muzerolle et~al. 2003; Kenyon et~al. 2005;
Burningham et~al. 2005; Caballero et~al. 2006, 2007).  
It is also the star forming region with the largest number of candidate isolated
planetary-mass objects with follow-up spectroscopy (Zapatero Osorio et~al. 2000,
2002b,c; Mart{\'\i}n et~al. 2001; Barrado~y~Navascu\'es et~al. 2001; Mart{\'\i}n
\& Zapatero Osorio~2003). Some of these planetary-mass objects
have discs (Zapatero Osorio et al. 2007; Scholz \& Eisl\"offel 2008;
Luhman et al. 2008). 

\begin{table*}
\caption[]{Identification number in Caballero (2006), 2MASS coordinates,
ASAS-3 $V$, DENIS $i$, and 2MASS $JHK_{\rm s}$ magnitudes of the 24 pre-selected 
stars.}  
\label{table.coo+phot+teff}
     $$ 
         \begin{tabular}{lccccccc}
            \hline
            \hline
            \noalign{\smallskip}
SO	& $\alpha$ 	& $\delta$	& $V$ 			& $i$ 			& $J$  			& $H$  			& $K_{\rm s}$  	          \\   
        & (J2000)	& (J2000)	& [mag]			& [mag]			& [mag]			& [mag]			& [mag]		          \\
            \noalign{\smallskip}
            \hline
            \noalign{\smallskip}
230838	& 05 37 36.86	& --02 08 17.7	&  $9.898\pm0.082$	&  $9.420\pm1.00$	&  $8.690\pm0.029$	&  $8.144\pm0.040$	&  $7.573\pm0.027$	   	  \\ 
000041  & 05 40 27.55	& --02 25 43.1	& $10.117\pm0.020$	&  $9.765\pm0.04$	&  $9.062\pm0.032$	&  $8.872\pm0.084$	&  $8.775\pm0.024$        \\ 
410305  & 05 40 12.09	& --03 05 28.4	& $11.035\pm0.031$	& $10.565\pm0.06$	&  $9.905\pm0.026$	&  $9.634\pm0.023$	&  $9.607\pm0.025$        \\ 
210734 	& 05 37 53.04	& --02 33 34.4	& $11.649\pm0.059$	& $10.798\pm1.00$	&  $9.991\pm0.027$	&  $9.605\pm0.024$	&  $9.474\pm0.025$	   \\ 
131161 	& 05 38 56.81	& --02 04 54.5	& $12.343\pm0.112$	& $11.147\pm0.04$	& $10.330\pm0.027$	&  $9.763\pm0.024$	&  $9.619\pm0.023$	   \\ 
420039  & 05 38 59.55	& --02 45 08.1	& $12.520\pm0.101$	& $11.686\pm0.03$	& $11.222\pm0.027$	& $10.984\pm0.023$	& $10.893\pm0.027$	   \\ 
210434 	& 05 38 07.85	& --02 31 31.4	& $12.715\pm0.112$	& $11.634\pm1.00$	& $10.566\pm0.027$	&  $9.930\pm0.023$	&  $9.769\pm0.025$	   \\ 
000005$^{a}$ 	& 05 38 38.23	& --02 36 38.4	& $12.9-14.0$	        & $12.340\pm0.03$	& $11.158\pm0.026$	& $10.465\pm0.023$	& $10.312\pm0.022$	   	 \\ 
211394  & 05 37 15.37	& --02 30 53.4	& $12.986\pm0.150$	& $12.176\pm1.00$	& $11.364\pm0.024$	& $11.006\pm0.025$	& $10.877\pm0.025$	   \\ 
420316 	& 05 38 54.11	& --02 49 29.8  & $13.164\pm0.204$	& $11.729\pm0.03$	& $10.829\pm0.026$	& $10.310\pm0.024$	& $10.126\pm0.019$	   \\ 
111112	& 05 38 48.04	& --02 27 14.2	& $13.340\pm0.341$	& $11.359\pm0.03$	& $10.156\pm0.023$	&  $9.463\pm0.026$	&  $9.187\pm0.019$        \\ 

130452	& 05 39 45.15	& --02 04 53.9	& $13.453\pm0.215$	& $12.236\pm0.04$	& $10.693\pm0.030$	&  $9.960\pm0.035$	&  $9.556\pm0.026$        \\ 

230062 	& 05 38 23.88	& --02 05 42.0	& $13.675\pm0.260$	& $12.378\pm0.03$	& $11.451\pm0.023$	& $10.866\pm0.026$	& $10.715\pm0.021$        \\ 
121137  & 05 38 35.87	& --02 30 43.3	& $13.701\pm0.290$	& $12.483\pm0.03$	& $11.245\pm0.026$	& $10.598\pm0.023$	& $10.424\pm0.024$        \\ 

000006$^{a}$  & 05 38 49.17	& --02 38 22.2	& $13.7-15.0$	        & $12.891\pm0.02$	& $11.389\pm0.026$	& $10.663\pm0.023$	& $10.511\pm0.022$	   	  \\ 
430847  & 05 39 32.57	& --02 39 44.0	& $13.756\pm0.202$	& $12.385\pm0.17$	& $10.820\pm0.027$	& $10.104\pm0.024$	&  $9.917\pm0.019$        \\ 
430136 	& 05 40 22.56	& --02 33 46.9	& $13.951\pm0.207$	& $12.473\pm0.04$	& $11.158\pm0.028$	& $10.533\pm0.023$	& $10.362\pm0.023$        \\ 

420040  & 05 38 27.26 	& --02 45 09.7	& $14.192\pm0.472$	& $12.847\pm0.02$	& $11.955\pm0.028$	& $10.792\pm0.026$	&  $9.944\pm0.028$        \\ 
440660  & 05 39 54.66	& --02 46 34.1	& $14.253\pm0.354$	& $12.757\pm0.03$	& $11.054\pm0.028$	& $10.251\pm0.024$	&  $9.832\pm0.024$        \\ 

121112  & 05 38 40.27	& --02 30 18.5	& $14.273\pm0.389$	& $12.839\pm0.02$	& $11.512\pm0.026$	& $10.763\pm0.023$	& $10.395\pm0.025$        \\ 
420147  & 05 38 52.01	& --02 46 43.7	& $14.301\pm0.305$	& $12.719\pm0.02$	& $11.518\pm0.026$	& $10.774\pm0.024$	& $10.421\pm0.021$        \\ 
430967  & 05 39 25.20	& --02 38 22.0 	& $14.344\pm0.341$	& $13.028\pm0.16$	& $11.307\pm0.031$	& $10.451\pm0.023$	& $10.002\pm0.023$        \\ 

140159  & 05 40 05.11	& --02 19 59.1	& $14.493\pm0.324$	& $12.950\pm0.03$	& $11.459\pm0.026$	& $10.767\pm0.024$	& $10.542\pm0.021$        \\ 
420742  & 05 38 39.82	& --02 56 46.2	& $\gtrsim 14.5$	& $12.778\pm0.02$	& $11.413\pm0.027$	& $10.744\pm0.023$	& $10.439\pm0.021$        \\ 
   \noalign{\smallskip}
   \hline
\end{tabular}
     $$ 
\begin{list}{}{}
\item[$^{a}$] For these stars we provide two values of the $V$
magnitude. The first number corresponds to the ASAS-3 $V$ magnitude
and other values from Wolk (1996; SO000005) and Sherry et al. (2004;
SO000006).  
\end{list}
\end{table*}

A metallicity detemination in the $\sigma$~Orionis cluster is becoming
essential. The derived distance and age of the cluster (from fits to
theoretical isochrones) depends on its metallicity, as recently shown
by Sherry et~al. (2008). At the age of the cluster ($\tau \sim$
3\,Ma), the protoplanetary discs are transiting from optically thick
to optically thin. Giant planets may be forming at this moment
(Lissauer 1993; Pollack et~al. 1996; Boss 1997; Ida \& Lin 2004).
Therefore, the metallicity {\bf could play an important} r\^ole in the
planetary formation in $\sigma$~Orionis (e.g. through the
planet-metallicity correlation -- Fischer \& Valenti 2005).

Cunha et~al. (1998) reported that the metallicity of the Orion OB1 association
as a whole is [Fe/H] $=-0.16\pm0.11$\footnote{[Fe/H]\,$=\log [N$(Fe)$/N$(H)]$-\log
[N$(Fe)$/N$(H)$]_\odot$.}, which is similar to other determinations
using different samples of stars (e.g., Cunha \& Lambert 1994); see a review of
abundance ratios and Galactic chemical evolution in McWilliam (1997).
However, there have been very few metallicity studies in the $\sigma$~Orionis
cluster.  
According to Cunha et~al. (2000) and references therein, the iron abundance of
the early-G star \object{HD~294297} (SO000041, Mayrit~1659068) is slightly subsolar
([Fe/H]$_{\rm HD~294297}$ = --0.18). Nonetheless, although the star is
young ($\log \epsilon(\mathrm{Li})= 2.56$\footnote{$\log 
\epsilon(\mathrm{Li})=\log[N$(Li)$/N$(H)]+12.}) and
follows the spectrophotometric sequence of $\sigma$~Orionis, its
cluster membership is in doubt because of its abnormally high proper
motion (Caballero 2007). 
Other star whose metallicity has been investigated with optical spectroscopy
is the K3 spectroscopic binary OriNTT~429~AB (Mayrit~1415279~AB).
Lee, Mart\'{\i}n \& Mathieu (1994) determined [Fe/H] = --0.43 and
--0.24$\pm$0.10 for the primary and the secondary, respectively.
A difference of iron abundances like that is questionable if none of the
components of the coeval system has not suffered yet from chemical
evolution (e.g. departure of main sequence).
The remaining metallicity study was presented by Caballero (2006), who
measured an average value of [Fe/H] = 0.0$\pm$0.1 from an ensemble of
mid-resolution spectra centred on H$\alpha$~$\lambda$6562.8\,\AA.
Using X-ray spectral energy distributions, coronal abundances of
cluster stars have also been found to be roughly solar
(Skinner et~al. 2008) or slightly lower (Sanz-Forcada, Franciosini \&
Pallavicini 2004; Franciosini, Pallavicini \& Sanz-Forcada 2006;
L\'opez-Santiago \& Caballero 2008). 

\begin{table*}
\caption[]{Names, IRFM effective temperatures, radial and rotational velocities, and
remarks from the literature of the 24 pre-selected stars.} 
\label{table.names+vel+remarks}
     $$ 
\begin{tabular}{lllcccl}
   \hline
   \hline
   \noalign{\smallskip}
SO$^a$	& Mayrit$^b$	& Simbad            	        & $T_{\rm eff}$       & $V_{r}^c$ & $v \sin i$& Remarks$^d$   \\ 
        &		& name            		& [K] 		      &      [\kms]  & [\kms]	 &		\\
            \noalign{\smallskip}
            \hline
            \noalign{\smallskip}
230838	& ...		& \object{HD 290772}           	& 6000$\pm$600$^{e}$  & $+23.3\pm0.4$ & $30\pm5$ & Herbig Ae/Be, {\em IRAS}, $\rho >$ 30\,arcmin \\
000041  & 1659068	& \object{HD 294297}           	& 6450$\pm$105        & $+24.6\pm0.2$ & $<20$	 & [Fe/H], $V_r$ = +25.0\,\kms \\
410305  & ...		& \object{HD 294308}           	& 6250$\pm$90	      & $-37.2\pm0.2$ & $<20$	 & $\rho >$ 30\,arcmin \\
210734  & 789281	& \object{2E 1454}     		& 5235$\pm$100        & $+24.0\pm0.3$ & $30\pm5$ & H$\alpha$ in broad emission \\
131161  & ...		& ...                   	& 4765$\pm$115        & $+23.4\pm0.3$ & $<20$	   & $\rho >$ 30\,arcmin, in NE nebulosity \\
420039  & 591158	& \object{[W96] 4771--0026}     & 5935$\pm$195        & $+33.7\pm0.3$ & $60\pm5$ & [S~{\sc ii}] and [N~{\sc ii}] in emission \\
210434  & 615296	& \object{2E 1459}     		& 4630$\pm$110        & $+24.5\pm0.4$ & $<20$	 & H$\alpha$ in faint emission \\
000005  & 105249	& \object{[W96] rJ053838--0236} & 4935--4000$^{f}$    & $+25.1\pm0.3$ & $<20$	 & H$\alpha$ in faint emission, $V_r$ = +42$\pm$10\,\kms \\
211394  & 1374283	& \object{Mayrit~1374283}       & 5300$\pm$210        & $+27.8\pm0.3$ & $<20$	 & high background around H$\alpha$ \\
420316  & 822170	& \object{RX J0538.9--0249}    	& 4520$\pm$175        & $+31.8\pm0.4$ & $<20$	 & H$\alpha$ in faint emission, $v \sin i$ = 28\,\kms \\
111112	& 528005 AB	& \object{[W96] 4771--899 AB}  	& $\lesssim$3500      & $+30.7\pm0.4$ & $<20$	 & classical T~Tau, resolved binary, \\ 
	& 		& 	  			&		      & 	      & 	 & $V_r$ = +31$\pm$10\,\kms \\
130452	& ...		& \object{RX J0539.8--0205 ABC} & $\lesssim$3800      & $+20.0\pm0.5$ & $60\pm5$ & H$\alpha$ in broad asymmetric emission, \\
	& 		& 	  			&		      & 	      & 	 & spectroscopic triple, $\rho >$ 30\,arcmin, \\
	& 		& 	  			&		      & 	      & 	 & in NE nebulosity, [S~{\sc ii}] and [N~{\sc ii}] in emission, \\
	& 		& 	  			&		      & 	      & 	 & $V_r$ = +27.8$\pm$0.9\,\kms \\
230062  & ...		& ...                   	& 4585$\pm$225        & $+22.6\pm0.3$ & $<20$	 & H$\alpha$ in very faint emission, $\rho >$ 30\,arcmin \\
121137  & 344337	& \object{2E 1468}    		& 4370$\pm$310        & $+28.9\pm0.4$ & $70\pm5$ & H$\alpha$ in broad asymmetric emission, \\ 
	& 		& 	  			&		      & 	      & 	 & $V_r$ = +35$\pm$7\,\kms, $v \sin i$ = 80$\pm$15\,\kms \\
000006  & 157155	& \object{[W96] rJ053849--0238} & 4450--3500$^{f}$    & $+34.1\pm0.7$ & $<20$	 & H$\alpha$ in faint, broad emission \\
430847  & 750107	& ...                   	& $\lesssim$3800      & $+30.7\pm0.5$ & $<20$	 & H$\alpha$ in faint emission \\
430136  & 1471085	& \object{Kiso A--0904 105}    	& $\lesssim$4000      & $+28.6\pm0.4$ & $60\pm5$ & H$\alpha$ in faint asymmetric emission, \\
	& 		& 	  			&		      & 	      & 	 & close to Horsehead, [S~{\sc ii}] and [N~{\sc ii}] in emission, \\
420040  & 609206	& \object{V505 Ori}            	& $\lesssim$3800      & $+29.5\pm0.7$ & $<20$	 & classical T~Tau, $V_r$ = +48$\pm$10\,\kms \\
440660  & 1223121	& \object{V606 Ori}            	& $\lesssim$3800      & $+29.3\pm0.4$ & $<20$	 & H$\alpha$ in strong, broad, asymmetric emission, \\
	& 		& 	  			&		      & 	      & 	 & class II, [N~{\sc ii}] in emission \\
121112  & 348349	& \object{Haro 5--13} 		& $\lesssim$3800      & $+34.3\pm0.4$ & $<20$	 & classical T~Tau \\
420147  & 653170	& \object{RU Ori}              	& $\lesssim$3800      & $+32.0\pm0.6$ & $<20$	 & classical T~Tau \\
430967  & 622103	& \object{BG Ori}              	& $\lesssim$3500      & $+27.0\pm0.6$ & $<20$	 & H$\alpha$ in strong, broad, asymmetric emission, \\
	& 		& 	  			&		      & 	      & 	 & class II, [S~{\sc ii}] in emission \\
140159  & 1541051	& \object{[NYS99] C--05}    	& $\lesssim$3500      & $+27.5\pm0.4$ & $<20$	 & H$\alpha$ in faint, broad emission \\
420742  & 1248183	& \object{[SWW2004] 125} 	& $\lesssim$3500      & $+29.3\pm0.5$ & $<20$	 & H$\alpha$ in faint emission \\
   \noalign{\smallskip}
   \hline
\end{tabular}
     $$ 
\begin{list}{}{}
\item[$^{a}$] Identification number in Caballero (2006). 
\item[$^{b}$] Mayrit designation in Caballero (2008c). 
\item[$^{c}$] Radial velocities extracted from cross-correlation with
a K3V template star (see main text).
\item[$^{d}$] Important remarks from the literature (see main text for
references).
\item[$^{e}$] Strong X-ray emitter (see
Section~\ref{effective.temperatures}).
\item[$^{f}$] Derived \teff from the two different $V$ magnitudes
given in Table~\ref{table.coo+phot+teff}. 
\end{list}
   \end{table*}

In this paper, we determine {\em photospheric} abundances of iron,
lithium and other elements in nine pre-main sequence, late-type stars
of $\sigma$~Orionis, with the aim of providing an average, accurate
value of its metallicity. Our study improves the [Fe/H] estimation
presented in Caballero (2006), from where we have taken the original
optical spectra. 

\section{Observations and data compilation}

Caballero (2006) presented the spectra of 143 stars in a wide area centred on
\object{$\sigma$~Ori~AB} (the OB multiple star system that
gives the name to the cluster).
Multi-object spectra were taken with the Wide Field Fibre Optical
Spectrograph (fiber diameter $\sim1.6$\,arcsec) and the robot
positioner AutoFib2 (WYFFOS+AF2) at the Observatorio del Roque de los
Muchachos, using the 4.2\,m {William Herschel Telescope} (WHT). 
We carried out the observations on 2003~Nov~30, covering the spectral
region $\lambda\lambda$6400--6800\,{\AA} at an effective resolving power
$\lambda / \Delta\lambda \sim 7500$ (or 0.83~{\AA}/pixel).     
The spectra were reduced, wavelength calibrated, and combined in a standard
manner using IRAF and the package {\tt dohydra}.  
Some preliminar results, like the frequency of accretors, were advanced by
Caballero (2005).
Further details on the observations, reduction, and analysis of the data and
discussion of the results will be presented in a forthcoming paper (Caballero
et~al., in~prep.~I)

Here, we picked up 25 stars in Caballero (2006) to investigate in detail
the metallicity in $\sigma$~Orionis.
Their spectra have the highest signal-to-noise ratio of the WYFFOS+AF2
sample and \ion{Li}{i}~$\lambda$6707.8\,{\AA} in manifest absorption, which
authenticates them as very young stars.
Furthermore, some of these stars have H$\alpha$~$\lambda$6562.8\,{\AA} in
strong, asymmetric, and/or broad emission, which is indicative of accretion
from a disc. 
In the full WYFFOS+AF2 sample, there exist stars with higher signal-to-noise
ratios, but they are too warm for our purposes (e.g. B2Vp
\object{$\sigma$~Ori~E} [Mayrit~42062]) or are bright GKM dwarfs in the
foreground (without trace of lithium; e.g. G0V \object{HD~294269}).

The determination of the optical magnitudes of \object{Mayrit~958292} was
affected by its close angular separation to the brighter star
TYC~4771--962--1 (\object{Mayrit~968292}).
Besides, the optical spectrum of Mayrit~958292 suffered from an incorrect
wavelength calibration in Caballero (2006).
As a result, we discarded the star from the input sample of pre-selected stars
suitable for our chemical analysis.

Identification number in Caballero (2006), coordinates from the Two Micron All
Sky Survey (2MASS -- Skrutskie et~al. 2006), $V$ magnitude from the All Sky
Automated Survey (ASAS-3 -- Pojmanski 2002), $i$ magnitude from the Deep Near
Infrared Survey from the Southern Sky (DENIS -- Epchtein et~al. 1997), and $J$,
$H$, and $K_{\rm s}$ magnitudes from 2MASS are given in
Table~\ref{table.coo+phot+teff} for each of the remaining 24 pre-selected stars.
The $V$ magnitudes in the ASAS-3 and Johnson systems are identical.
Tabulated $V$ values were computed by averaging all ASAS-3 measurements at less
than 15\,arcsec to the central 2MASS coordinates of each star and with ASAS-3
photometric quality flags ``A'' and ``B''. 
The error $\delta V$ is the standard deviation of the mean $V$.
Large $\delta V$ values indicate both low signal-to-noise ratio and intrinsic
photometric variability.
For example, the binary SO111112 is one of the most variable stars in the Orion
Belt according to Caballero et~al. (in~prep.~II).
The $V$ magnitude of SO420742 is an approximate lower limit (Sherry
et~al. [2004] measured $V \approx$ 14.72\,mag).
Likewise, in Table~\ref{table.names+vel+remarks}, we provide for each star (when
available) its Mayrit designation, Simbad name, IRFM effective
temperatures, radial and rotational
velocities, and important
remarks from the literature and from our spectra. 
Many of the remarks have been taken from the Mayrit catalogue (Caballero 2008c), 
although we also provide additional data (e.g. $V_r$, $v \sin i$, type of
H$\alpha$ emission) mostly from Zapatero Osorio et~al. (2002a) and
Caballero (2006). See below for details on the computation of the
radial and rotational velocities.

\section{Analysis and results}

\subsection{Effective temperatures}
\label{effective.temperatures}

Effective temperatures of the 24 pre-selected stars, \teffo, were
determined using the infrared flux method (IRFM) and the magnitudes $V$, $J$,
$H$ and $K_{\rm s}$ (see Table~\ref{table.names+vel+remarks}).
We used the calibration of the bolometric flux as a function of
the colour $V-K_{\rm s}$ and the magnitude $K_{\rm s}$ from Alonso
et~al. (1995, 1999), converted into the homogeneous system of Bessell \& Brett (1988).
Each of these temperatures was determined by comparing the theoretical fluxes
integrated in the 2MASS filters with the observed $JHK_{\rm s}$ magnitudes. We
assumed solar metallicity and surface gravity \logg $=3.9$ for all the stars in 
the sample. Systematic errors due to a different metallicity (by +0.1 dex) and a
different surface gravity (by +0.5 dex) are in the range $7-100$\,K and
$2-30$\,K  respectively, for all stars in the sample.
The derived effective temperature was the average of three individual
\teffo, one for each near-infrared passband, weighted
with the inverse of their individual errors, one for each near-infrared passband. 
To estimate the uncertainties in \teffo, we considered a linear transmission of the 
errors, given that the errors in each photometric magnitude are not totally 
independent. 
The error in the average effective temperature accounts for the photometric
errors of the observed $JHK_{\rm s}$ and $V$ magnitudes, the error on the absolute
calibration of 2MASS (Cohen et~al. 2003), uncertainties on surface gravity
($\Delta_{\log{g}}=0.5$ dex) and metallicity 
($\Delta_{[{\rm Fe}/{\rm H}]}=0.1$ dex).  
Further details will be provided by Gonz\'alez Hern\'andez \&
Bonifacio (in~prep.).
 
A colour excess of $E(B-V)=0.082$\,mag towards the $\sigma$~Orionis cluster was
derived from the maps of dust infrared emission of Schlegel et~al. (1998).
This colour excess is consistent with the range of colour excess
estimated by Sherry et~al. (2008) for all early-type star members of
the $\sigma$~Orionis cluster, 0.04\,mag $< E(B-V) <$ 0.09\,mag.

Some of the values might be slightly affected by flux excesses at the
near-infrared bands because of circumstellar discs, which make derived
$T_{\rm eff}$ to be higher than in young disc-less stars of the same
$V$ magnitude. However, we do not see any significant flux excess in
our stars except for the Herbig Ae/Be star SO230838 (HD 290772;
Gregorio-Hetem \& Hetem 2002). It is an intense accretor, {\em IRAS}
source, X-ray emitter, F6-type star much warmer than tabulated in
Table~\ref{table.names+vel+remarks}. For this star, IRFM temperatures
derived from the three 2MASS filters differ by 600, while for other
stars in the sample the three IRFM temperatures are usually 
consistent within 100\,K.
Besides, for another two stars, S0000005 and SO000006, two values of
the $V$ magnitude are given, providing two temperature determination
(see the discussion in Section~\ref{spectra.selection}).

\subsection{Final sample for chemical analysis}
\label{spectra.selection}

Several of the 24 pre-selected stars are not appropiate for the
chemical analysis. 
In particular, there are 11 stars that have \teff $\lesssim$
4000\,K, for which the IRFM temperature estimates are not reliable
because the calibration of the bolometric flux is uncertain at these
temperatures. 
In addition, our spectral fitting procedure is not so accurate at these low
temperatures because we do not include molecular bands, in particular, TiO
bands, which appear to be completely necessary for the spectral fitting at
\teff $\lesssim 4200$\,K.   
Among the discarded ``cool'' stars, one is a close binary resolved with
adaptive optics observations (SO111112; $\rho$ = 0.40$\pm$0.08\,arcsec --
Caballero 2005) and another one is a spectroscopic triple with a high apparent
rotational velocity (SO130452 -- Alcal\'a et~al. 2000).
Both of them likely harbour discs because they display
flux excesses in the near-infrared.

The stars S0000005 and SO000006 are two X-ray emitters with very
strong \ion{Li}{i} $\lambda$6708\,{\AA} lines. The ASAS-3 $V$
magnitudes provide effective temperatures derived from IRFM (\teff =
4935 and 4450\,K, respectively) that must be too high because they
produces too large lithium abundances ($\log \epsilon(\mathrm{Li})
\sim 4.4$). Other $V$ magnitudes tabulated in the literature for both
stars (Zapatero Osorio et al.2002a; Sherry et al. 2004) give IRFM
effective temperature below 4000\,K. 

In order to perform a detailed chemical analysis, we discarded 
the 13 ``cool'' stars, including S0000005 and SO000006, the very
active Herbig Ae/Be star SO230838, and the 
possible spectroscopic binary SO410305 (Section~\ref{radial.velocity})
and retained only nine stars appropiate for spectral~fitting.  

According to Siess et~al. (2000) models for very young ages, the
derived effective temperatures of the nine selected stars correspond to
approximate spectral types in the interval F5 to K6.

\begin{figure*}
\centering
\includegraphics[width=0.49\textwidth]{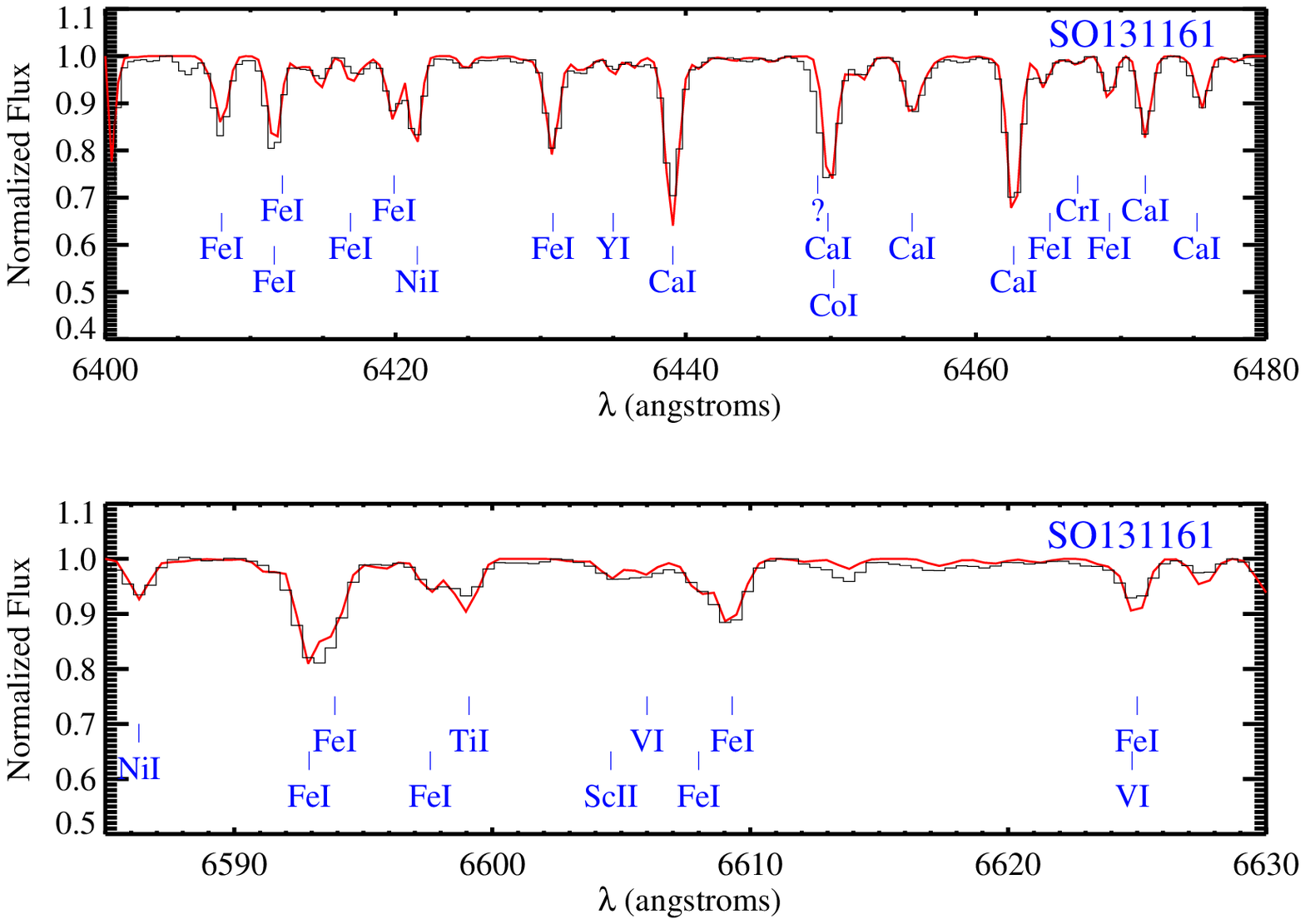}
\includegraphics[width=0.49\textwidth]{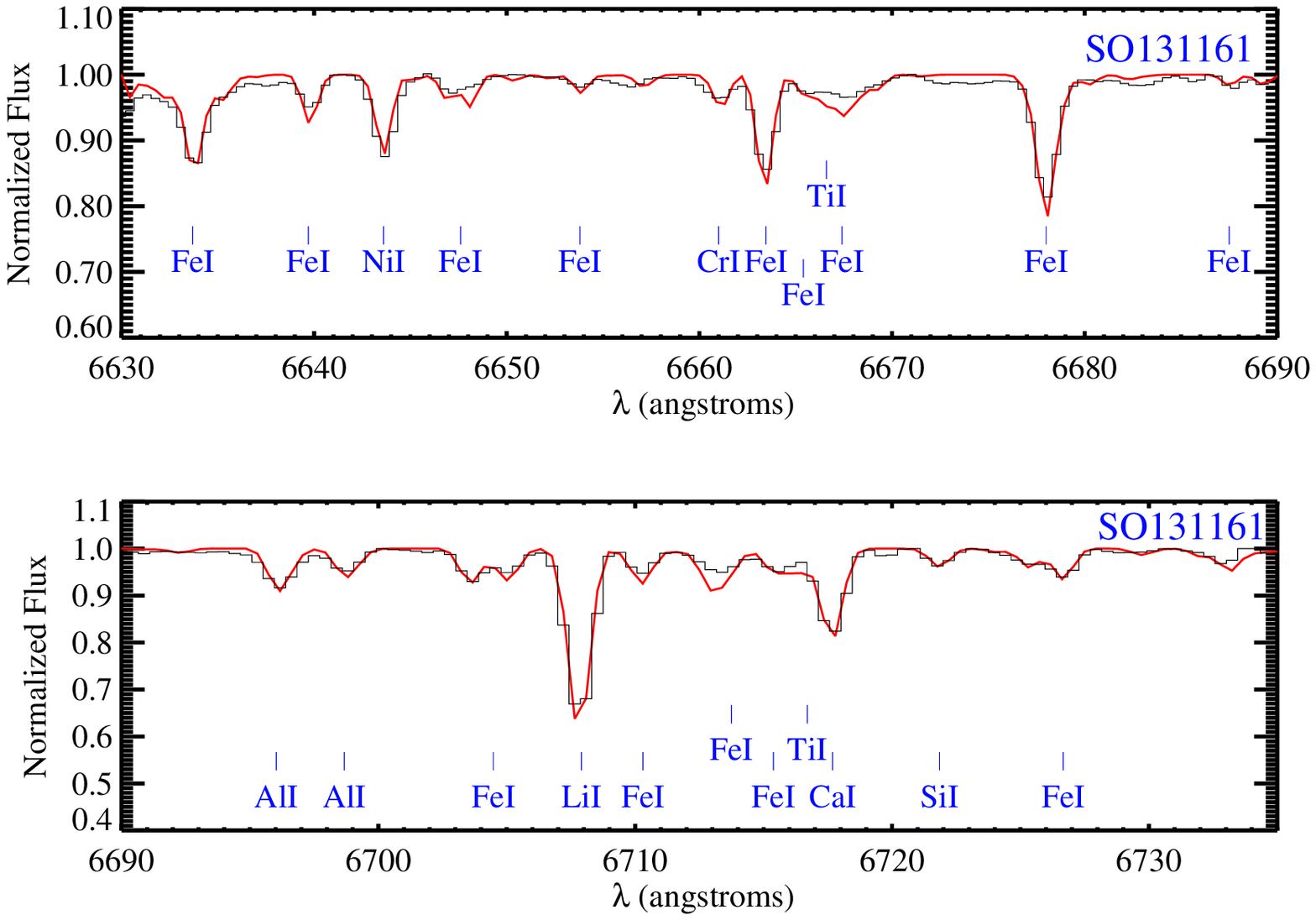}
\caption{WYFFOS+AF2 spectrum of SO131161 at the four investigated spectral
intervals in comparison with synthetic spectra computed for the element
abundances shown in Table~\ref{table.stellarparameters}.
Some absorption features are marked.} 
\label{fig2}
\end{figure*}  

\subsection{Radial and rotational velocities}
\label{radial.velocity}

For all stars in the pre-selected sample (including the nine stars
suitable for chemical analysis), we extracted their radial velocities by
cross-correlating their WYFFOS+AF2 spectra with the spectrum of a
K3V template star taken from the S$^4$N database (\object{HD~160346} [HIP 86400]
-- Allende Prieto et~al. 2004), using the 
software {\scshape MOLLY} developed by T.~R.~Marsh. 
We also computed the optimal rotational velocity, $v \sin{i}$, by subtracting
broadened versions of the template star, in steps of 5\,\kmso, and minimizing
the residual.  
We used an spherical rotational profile with linearized limb-darkening $\epsilon
= 0.65$ (Al-Naimiy 1978). 
We took into account the instrumental broadening, which was estimated at about
$40$\,\kmso. Radial and rotational velocities are provided in
Table~\ref{table.names+vel+remarks}. 

All except one of the stellar radial velocities given in
Table~\ref{table.names+vel+remarks} are in agreement with the canonical value of
the average cluster radial velocity at $V_r \approx$ +30.9$\pm$0.9\,\kms
(Wilson 1953; Morrell \& Levato 1991; Zapatero Osorio et~al. 2002a; Kenyon et~al.
2005; Maxted et~al. 2008; Sacco et~al. 2008). 
The outlier, a late F-star with lithium (\teff = 6250$\pm$90\,K, $V_r$ =
--34$\pm$3\kmso; SO410305), could be a spectroscopic binary.
Indeed, the star with the second most discordant $V_r$ is the spectroscopic
triple SO130452 ($V_r$ = +20.0$\pm$0.5\kmso).

We have measured an average radial velocitie $\overline{V_{r}}$ =
+27.9$\pm$4.0\,\kms ~($N$ = 23), where the error is the standard
deviation of the mean. 
Since the average error in the individual measurements of the
cross-correlation method is $\overline{\delta V_{r}}$ = 0.4\,\kmso,
the simplest interpretation is {\bf the $\sigma_{V_r} \sim$ 4\,\kms
would be the cluster velocity dispersion.} However, other authors (e.g.
Sacco et~al. 2008) have derived lower velocity dispersions, {\bf at
$\sigma_{V_r} \sim$ 1\,\kmso}. This value is of fundamental
importance for some theoretical scenarios but, unfortunately, has not
been well treated in the literature of the $\sigma$~Orionis cluster.

\subsection{Metallicities and gravities}
\label{metallicities.gravities}

\begin{table*}
\caption[]{Basic stellar parameters of the nine selected stars.}  
\label{table.stellarparameters}
     $$ 
\begin{tabular}{lccccccccc}
\hline
\hline
\noalign{\smallskip}
SO 	& \teff 	& \logg 	& EW(\ion{Li}{i})$^{a}$  & 
$[{\rm Fe}/{\rm H}]$ & $[{\rm Al}/{\rm H}]$ & $[{\rm Si}/{\rm H}]$ &
$[{\rm Ca}/{\rm H}]$ & $[{\rm Ni}/{\rm H}]$  & $\log
\epsilon(\mathrm{Li})_\mathrm{NLTE}$   \\       
        & [K]		&             	& [{\AA}] & & & & & & \\
            \noalign{\smallskip}
            \hline
            \noalign{\smallskip}
000041 	& 6450$\pm$105 	& $3.4\pm0.3$   & $0.04\pm0.01$ & --0.09 $\pm$ 0.10   &   0.10 $\pm$ 0.12  &  0.10 $\pm$ 0.13  & --0.12 $\pm$ 0.13  & --0.05 $\pm$ 0.12  & 2.75 $\pm$ 0.18  \\
210734 	& 5235$\pm$100 	& $3.8\pm0.3$   & $0.29\pm0.01$ & --0.02 $\pm$ 0.10   & --0.12 $\pm$ 0.12  &  0.20 $\pm$ 0.13  &   0.03 $\pm$ 0.11  & --0.12 $\pm$ 0.14  & 3.07 $\pm$ 0.18  \\
131161 	& 4765$\pm$115 	& $3.9\pm0.3$   & $0.42\pm0.01$ &   0.13 $\pm$ 0.10   &   0.03 $\pm$ 0.12  &  0.23 $\pm$ 0.13  &   0.18 $\pm$ 0.13  &	0.18 $\pm$ 0.14  & 3.43 $\pm$ 0.18  \\
420039 	& 5935$\pm$195  & $3.7\pm0.3$   & $\le  0.14^b$ & --0.10 $\pm$ 0.15   &   0.15 $\pm$ 0.14  &  0.00 $\pm$ 0.15  & --0.22 $\pm$ 0.21  & --0.10 $\pm$ 0.13  & 2.93 $\pm$ 0.31  \\
210434 	& 4630$\pm$110 	& $3.8\pm0.3$   & $0.45\pm0.01$ &   0.08 $\pm$ 0.10   & --0.04 $\pm$ 0.12  &  0.23 $\pm$ 0.13  &   0.15 $\pm$ 0.11  &	0.15 $\pm$ 0.14  & 3.70 $\pm$ 0.18  \\
211394 	& 5300$\pm$210  & $4.0\pm0.3$   & $0.22\pm0.01$ & --0.15 $\pm$ 0.10   & --0.05 $\pm$ 0.14  &  0.00 $\pm$ 0.15  & --0.20 $\pm$ 0.21  & --0.10 $\pm$ 0.12  & 2.77 $\pm$ 0.31  \\
420316 	& 4520$\pm$175 	& $4.0\pm0.3$   & $0.44\pm0.01$ & --0.08 $\pm$ 0.13   & --0.15 $\pm$ 0.14  &  0.20 $\pm$ 0.15  & --0.15 $\pm$ 0.21  &	0.07 $\pm$ 0.16  & 3.16 $\pm$ 0.31  \\
230062 	& 4585$\pm$225  & $4.1\pm0.3$   & $0.49\pm0.01$ &   0.02 $\pm$ 0.17   & --0.10 $\pm$ 0.14  &  0.10 $\pm$ 0.15  & --0.07 $\pm$ 0.21  &	0.10 $\pm$ 0.19  & 3.61 $\pm$ 0.31  \\
121137 	& 4370$\pm$310	& $3.9\pm0.3$	& $\le  0.47^b$ &   0.00 $\pm$ 0.20   & --0.05 $\pm$ 0.17  &  0.05 $\pm$ 0.19  & --0.20 $\pm$ 0.31  &	0.07 $\pm$ 0.16  & 3.25 $\pm$ 0.45  \\
\noalign{\smallskip}
\hline     
\noalign{\smallskip}
Cluster$^{\rm c}$ & ... & ... & ... & $-0.02\pm0.09$ & $-0.03\pm0.10$ & $ 0.12\pm0.09$ & $-0.07\pm0.15$  & $ 0.02\pm0.12$ &  ... \\
\noalign{\smallskip}
\hline
\end{tabular}
     $$ 
\begin{list}{}{}
\item[$^{a}$] Equivalent width of \ion{Li}{i} $\lambda$6708\,{\AA}.
\item[$^{b}$] The lithium line is blended with the \ion{Fe}{i} lines at 6703--5 {\AA} 
due to high rotational velocity of the star.
\item[$^{c}$] Average abundances of the $\sigma$~Orionis cluster.
Uncertainties are computed from standard deviations of the
measurements from each star.
\end{list}
\end{table*}

Using the derived effective temperatures, we tried to infer the
surface gravity, \loggo, and the metallicity, [Fe/H], of the nine
selected stars, using a minimization routine that compares several features of the stellar
spectra with a grid of synthetic spectra (see Gonz\'alez Hern\'andez
et~al. 2004, 2005, 2008). 
The synthetic spectra were computed using the local thermodinamic equilibrium
(LTE) code MOOG (Sneden 1973), a grid of LTE model
atmospheres (Kurucz 1993), and atomic line data extracted from the Vienna Atomic
Line Database (VALD; Piskunov 1995).  
The oscillator strengths of relevant lines were adjusted until they
reproduced the solar atlas (Kurucz et~al. 1984) with solar element abundances
(Grevesse et~al.~1996).  
The adopted abundances were:
$\log \epsilon(\mathrm{Al})_{\odot}=6.47$,  
$\log \epsilon(\mathrm{Si})_{\odot}=7.55$,  
$\log \epsilon(\mathrm{Ca})_{\odot}=6.36$,
$\log \epsilon(\mathrm{Fe})_{\odot}=7.50$, and  
$\log \epsilon(\mathrm{Ni})_{\odot}=6.25$.
Element abundances (calculated assuming LTE) were computed through the
expression $\mathrm{[X/H]}= \log [N(\mathrm{X})/N(\mathrm{H})]_{\rm star} - \log
[N(\mathrm{X})/N(\mathrm{H})]_\odot$, where $N(\mathrm{X})$ is
the number density of~atoms

We selected 12 spectral features containing in total 31 lines of
\ion{Fe}{i} and, in some cases, six lines of \ion{Ca}{i}, with 
excitation potentials between 1 and 5\,eV. 
For each given iron abundance in the interval $[{\rm
Fe}/{\rm H}] < 0$, the calcium abundance was fixed according to the
Galactic trend of calcium (Bensby et~al.\ 2005), while for $[{\rm
Fe}/{\rm H}] > 0$, we assumed $[{\rm Ca}/{\rm Fe}]=0$. 

Microturbulence, $\xi$, was fixed in each atmospheric model according
to the calibration as a function of effective temperature and surface
gravity. 
Such calibration has been derived for stars of the solar neighbourhood
with approximate solar metallicity (Allende Prieto et~al. 2004).  

We imposed the surface gravity and the metallicity to range, in steps of
0.05\,dex, in the intervals $2.50 <\log{g}< 4.50$ and $-0.50 < [{\rm Fe}/{\rm
H}] < 0.50$, respectively.  
The comparison of this grid with the observed spectra, using a bootstrap
Monte~Carlo method for each \teffo, gives the most likely values of $\log{g}$
and $[{\rm Fe}/{\rm H}]$.
The values of the trio $T_{\rm eff}$-$\log{g}$-$[{\rm Fe}/{\rm H}]$ are provided
in Table~\ref{table.stellarparameters}. 
All the stars show a metallicity in approximate agreement with
the solar value within the error bars.

Our spectral-fitting technique also allows us to determine the veiling, defined
as $F_{\rm disc}/F_{\rm star}$, where $F_{\rm disc}$ and $F_{\rm star}$ are the
flux contributions of the disc and the continuum of the star, respectively. 
We have not found any clear evidence of veiling in the analysed spectral region
($\lambda\lambda$6400--6800\,{\AA}) of the
nine selected stars.  

\begin{figure*}
\centering
\includegraphics[width=0.49\textwidth]{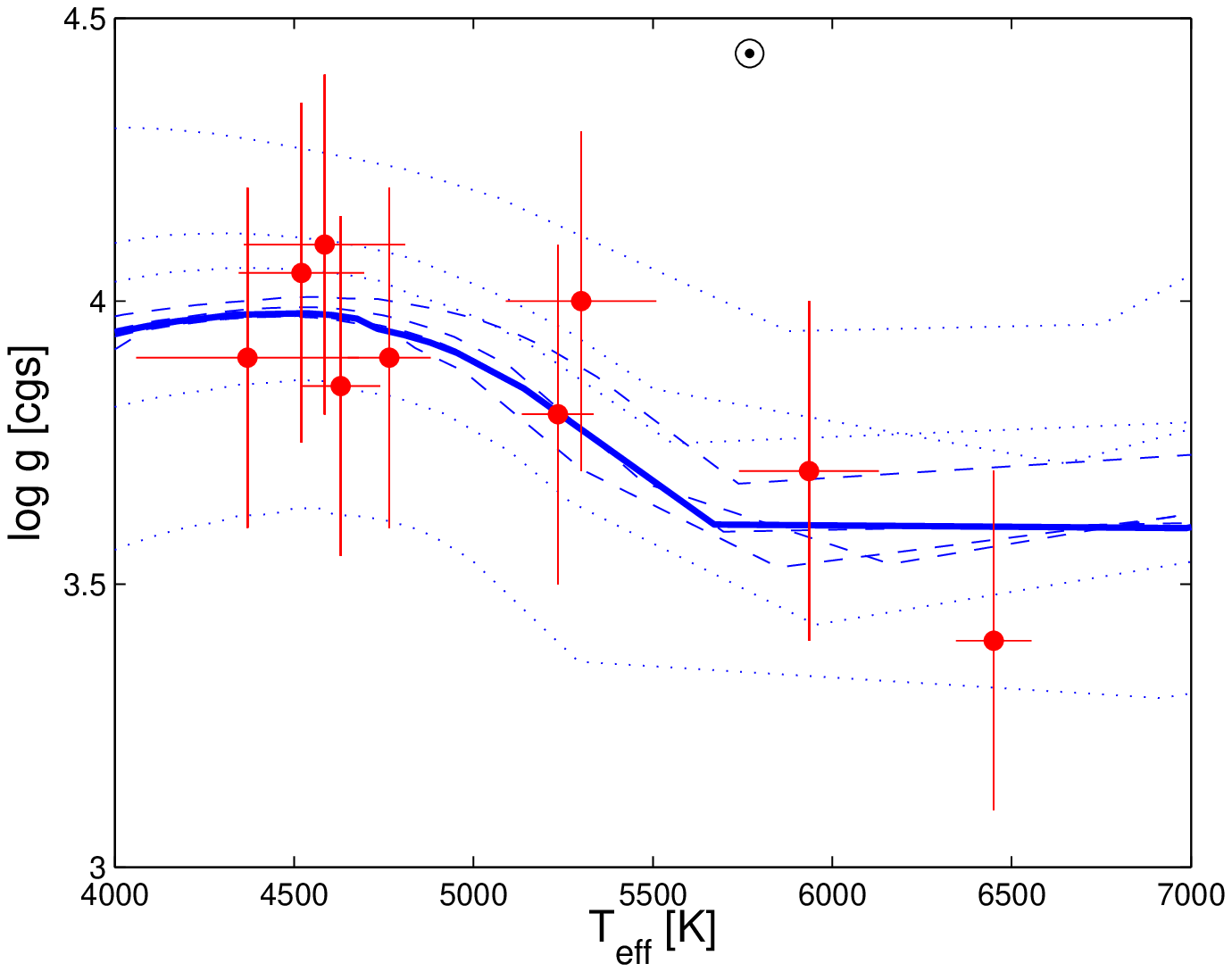}
\includegraphics[width=0.49\textwidth]{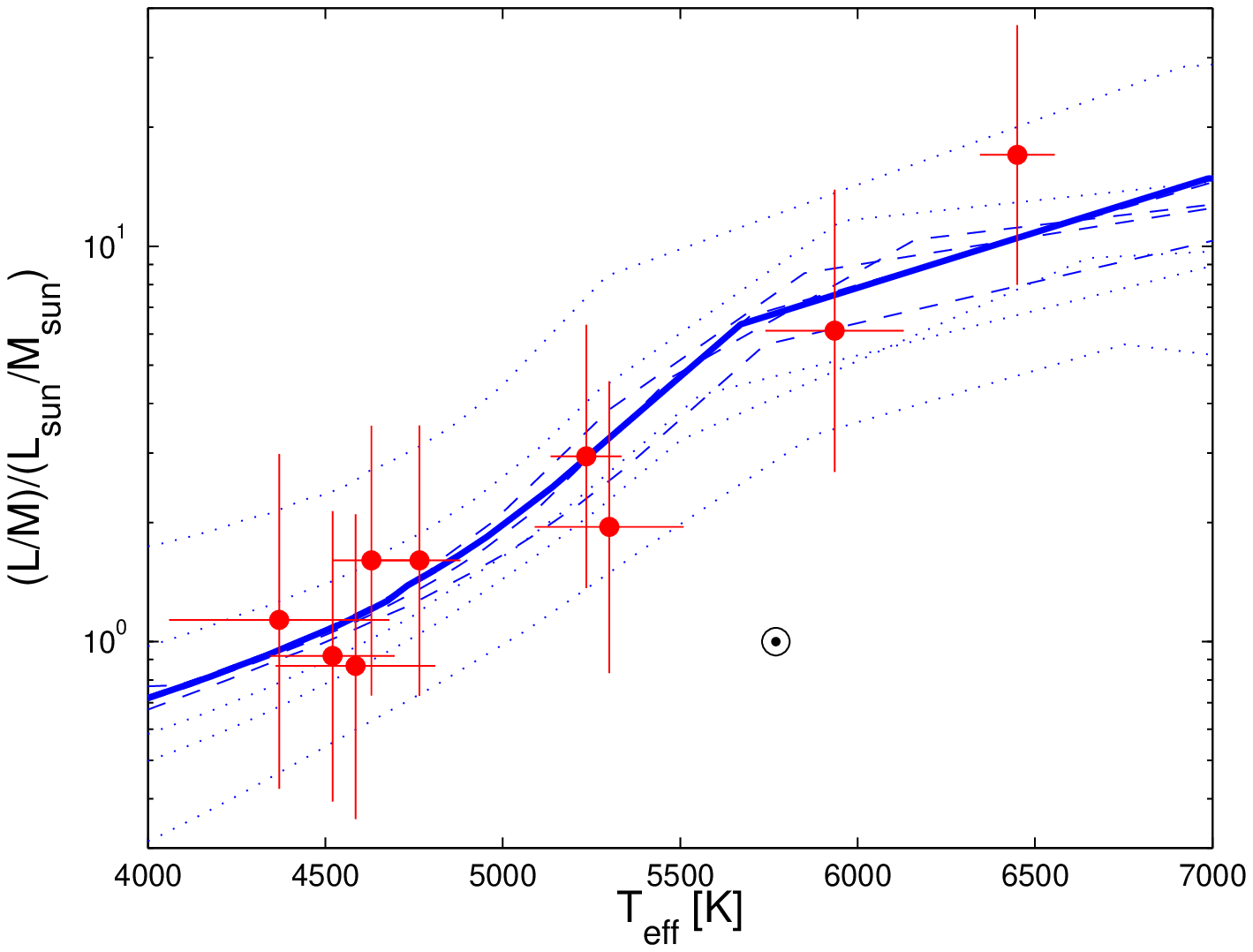}
\caption{Gravity ($\log{g}$, {\em left}) and luminosity per unit of mass
normalized to the Sun ($\mathcal{L}$, {\em right}) as functions of effective
temperature ($T_{\rm eff}$) of the nine selected stars (filled circles with
error bars).  
A bunch of Siess et~al. (2000) theoretical evolutionary tracks for
different ages, metallicities and overshooting are plotted. 
Dotted lines: tracks for $Z$ = 0.02, overshooting, and 10, 5, 4, 2,
and 1$\,$Ma, from top to bottom;  
dashed lines: tracks for 3$\,$Ma, {\em no} overshooting, and $Z$ = 0.01, 0.03,
and 0.04, from top to bottom;  
and thick solid line: track for 3$\,$Ma, $Z$ = 0.02, and overshooting.
The solar symbol, $\odot$, indicates the position of the Sun.} 
\label{fig4}
\end{figure*}

\subsection{Stellar abundances}
\label{section.stellarabundances}

Adopting the stellar parameters given in
Table~\ref{table.stellarparameters}, we determined the abundances of
aluminium, calcium, and nickel from several features of each element. 
The silicon and lithium abundances were derived from only one spectral
feature\footnote{The silicon feature is Si~{\sc i} $\lambda$6721.8\,\AA.}.
In Fig.~\ref{fig2}, we display all spectral regions analysed in this
work for a representative star of the selected sample.  
The element abundances are presented in Table~\ref{table.stellarparameters}. 
The errors on the element abundances show their sensitivity to the uncertainties
on the effective temperature, $\Delta_{T_{\mathrm{eff}}}$, surface gravity, 
$\Delta_{\log{g}}$, and the dispersion of the measurements from
different spectral features, $\Delta_{\sigma}$. 
The errors $\Delta_{\sigma}$ were estimated as $\Delta_{\sigma}
=\sigma/{N}^{1/2}$, where $N$ is the number of features analysed of each element
and $\sigma$ is the standard deviation of the $N$ measurements.
The errors $\Delta_{T_{\mathrm{eff}}}$ and $\Delta_{\log{g}}$ were determined as
$\Delta_{T_{\rm eff}}=(\sum_{i=1}^N\;\Delta_{T_{\rm eff},i})/N$ and
$\Delta_{\log{g}}=(\sum_{i=1}^N\;\Delta_{T_{\log{g}},i})/N$. 
For silicon and lithium abundances, the error associated to the
dispersion of the measurements, $\sigma$, was 
assumed to be the average dispersion of iron, calcium, and nickel
abundances (i.e. $\Delta_{\sigma} \equiv \overline{\sigma}$).
The total error given in Table~\ref{table.stellarparameters} was derived using the 
expresion $\Delta_{\rm tot}^2 = \Delta_{\sigma}^2 +
\Delta_{T_{\mathrm{eff}}}^2 + \Delta_{\log{g}}^2$.

All the elements abundances were derived with the assumption of local
thermodynamical equilibrium, except for lithium, for which we 
provided its non-equilibrium (non-LTE) abundance. 
We estimated the non-LTE abundance correction for this element from
the theoretical LTE and non-LTE curves of  growth in Pavlenko \&
Magazz\`u (1996). 

\section{Discussion}

\subsection{Stellar parameters and heliocentric distance}
\label{distance}

We display in Fig.~\ref{fig4} two diagrams involving basic stellar
parameters of the selected stars in comparison with several
theoretical evolutionary tracks taken from Siess et~al. (2000), for
several ages and metallicities. The derived stellar parameters are
consistent with almost all the theoretical tracks due to the large
error on surface gravity. However, their central values are very well
reproduced by the solid line that represents the 3\,Ma-old,
overshooting track. Ages lower than 2\,Ma or larger than 4\,Ma are
marginally consistent with the stellar parameters of the selected
stars. The theoretical tracks are only sensitive to
metallicity for effective temperatures higher than \teff $\sim$
4800\,K.  

In the right panel of Fig.~\ref{fig4}, we plot the luminosity per unit
of mass, normalized to this quantity in the Sun:  

\begin{equation}
\mathcal{L} = (L/M)/(L_\odot / M_\odot). 
\end{equation}

\noindent Using the well known expressions for the bolometric luminosity and the
surface gravity, it is derived~that: 

\begin{equation}
L/M = 4 \pi G \sigma \frac{T_{\rm eff}^4}{g},
\end{equation}

\noindent and, therefore,

\begin{equation}
\mathcal{L} = \frac{g_\odot}{g} \left( \frac{T_{\rm
eff}}{T_{{\rm eff,}\odot}} \right)^4.
\end{equation}

\noindent In another way, $\mathcal{L} \propto g^{-1} T_{\rm eff}^4$ at a given
metallicity, which means that the normalized luminosity per unit of mass only
depends on the derived stellar parameters.

In addition, in Fig.~\ref{fig6} we display the $\mathcal{L}$ vs. $J$
diagram and several theoretical evolutionary tracks for different 
heliocentric distances to the $\sigma$~Orionis cluster.  
In this case, all selected stars, except three, are
close to the tracks for distances in the range 350\,pc $< d <$
500\,pc.
These three stars are the hottest stars in the sample (SO000041 --\teff =
6450$\pm$105\,K--, SO420039 --\teff =
5935$\pm$195\,K-- and SO211394 --\teff = 5300$\pm$210\,K--).
SO000041 is marginally consistent with the track at $d=$ 400\,pc but still too
luminous for its expected position in this diagram (see
Section~\ref{lithium}). 
For the other two stars, one should expect $J$ magnitudes in the range 9\,mag
$\lesssim J \lesssim$ 10\,mag, but surprinsingly they have $J \sim$ 11.3\,mag. 
Remarkably, one of these two stars has a large rotational velocity ($v \sin{i}$ =
60$\pm$5\,\kmso; SO420039), while the another star falls in the halo of the
$\sigma$~Orionis cluster ($\rho \sim$ 23\,arcmin; SO211394), where the
contamination by neighbouring young stellar populations gets higher.
In any case, an heliocentric distance of $d \sim$ 400\,pc is in agreement with
recent results on the distance to the $\sigma$~Orionis cluster in Sherry et~al.
(2008) and Caballero (2008a).  
These authors claimed that distances of $d=420 \pm 30$ and $\sim$ 385\,pc,
respectively, are more plausible than previous estimates at~350\,pc.  
However, our error bars are so large that we cannot favour a distance
of 450\,pc rather than~350\,pc.

\subsection{Lithium abundance}
\label{lithium}

\begin{figure}
\centering
\includegraphics[width=0.49\textwidth]{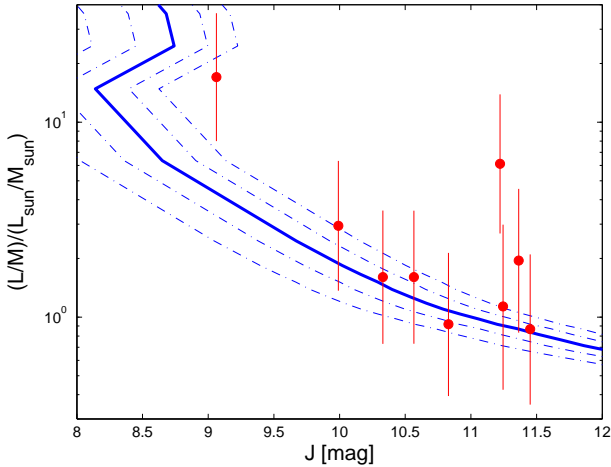}
\caption{Same as Fig.~\ref{fig4} but for the normalized luminosity per unit of
mass, $\mathcal{L}$, as a function of the apparent magnitude, $J$.
Dashed-dotted lines: tracks for fixed age (3\,Ma), metallicity ($Z$ = 0.02), and
overshooting, and variable heliocentric distances $d$ = 500, 450, 350,
300$\,$pc, from top to bottom; 
and thick solid line: track for 400$\,$pc, 3\,Ma, $Z$ = 0.02, and overshooting.
} 
\label{fig6}
\end{figure}

Five stars in the selected sample have
lithium abundances consistent within the error bars or slightly lower than the
solar meteoritic value, $\log \epsilon(\mathrm{Li}) = 3.3$, which is also
considered as the current Galactic or ``cosmic'' $^7$Li abundance
(Boesgaard \& Tripicco 1986).  
Therefore, young, population~I stars are believed to have inherited this
lithium abundance at birth.  
However, two stars (SO131161 and SO210434) show 
higher lithium abundances, $\log \epsilon({\rm Li}) \sim
3.6-3.7$ which might be due to wrong stellar parameters. 
Note that a 100\,K cooler effective temperature would provide a
0.10--0.15\,dex lower lithium abundance. 
We should note that the uncertainties on the
lithium abundances are quite high due to the large error bars of the
temperature determination. Thus, some stars have lithium abundances
that are at least marginally consistent, within the  
error bars, with the ``cosmic'' $^7$Li abundance.

On the other hand, such high lithium abundances have been found in
other T~Tauri stars. 
For instance, Mart{\'\i}n et~al. (1994) showed two stars with similar
stellar parameters (\teff $\sim 4700$\,K and \logg $\sim 3.6-4.0$) and LTE
lithium abundances $\log \epsilon({\rm Li}) \sim 3.65 - 3.60$. 
However, their NLTE corrections were between --0.3 and --0.2\,dex,
thus significantly larger than our NLTE corrections,
$\Delta_\mathrm{NLTE-LTE} \lesssim -0.10$\,dex. 
Those corrections moved their stars to the region of normal lithium
abundances. In addition, a high NLTE lithium abundance has also been
found in a double-lined spectroscopic binary of the $\sigma$~Orionis
cluster by Lee et~al. (1994): Mayrit~1415279~AB (it was presented in 
Section~\ref{introduction}). 

Likewise, there are two additional stars with a lithium abundance
slightly lower than the others, SO211394
($\log \epsilon({\rm Li})$ = 2.77$\pm$0.31) and 
SO000041 ($\log \epsilon({\rm Li})$ = 2.75$\pm$0.18).
SO211394 is one of the three outliers in the $\mathcal{L}$ vs. $J$
diagram in Fig.~\ref{fig6}.
The location of this star in the halo of the $\sigma$~Orionis cluster,
its relatively low lithium abundance, and the abnormally faint
$J$-band magnitude for its luminosity per unit of mass may suggest
classifying the star as a member of a neighbouring, not-so-young
population in the \object{Ori~OB1b} association different from
$\sigma$~Orionis (see, e.g., Jeffries et~al. 2006). Given its position
to the west of the cluster, SO211394 may belong to the extended
population of the $\sim$5\,Ma-old surrounding \object{Alnilam}
(\object{Collinder~70} or the ``$\epsilon$~Orionis cluster'';
Caballero \& Solano 2008). 
Nevertheless, its position in a EW(Li~{\sc i}) vs. \teff diagram (e.g.
L\'opez-Santiago et~al. 2006) might also indicate that the star
is a member of a distant ($d = 710 \pm 50$\,pc) moving group with a
Pleiades-like age. 

SO000041 has a position only marginally consistent with the 
theoretical tracks in the $\mathcal{L}$ vs. $J$ diagram. 
This fact, together with the rather high proper motion of
SO000041 (Section~\ref{introduction}) and its location even further from the
cluster centre than SO211394, bring doubts about its membership to the
$\sigma$~Orionis cluster.  
SO000041 is the easternmost, bright, cluster member candidate
in the Mayrit catalogue, just in the outskirts of the
\object{Horsehead Nebula} (Caballero 2007, 2008c). 
Membership in a slightly younger population of recently born stars
associated to the nebula might explain its relatively low gravity.
However, membership in a $\sim$100\,Ma-old moving group (just like SO211394) at
$d \sim 630 \pm 50$\,pc could explain both its relatively low lithium abundance
and large proper motion. Further astrometric and spectrophotometric
studies are needed to ascertain the real nature of SO000041 and SO211394.

\subsection{Metallicity of $\sigma$~Orionis}

The nine selected stars share the same radial velocity and have high
Li abundances consistent with the ``cosmic'' $^7$Li abundance, so we
think that they are probably members of the $\sigma$~Orionis cluster.
According to the values shown in Table~\ref{table.stellarparameters}, 
the average metallicity of these nine selected stars (i.e. the
metallicity of the $\sigma$~Orionis cluster) is [Fe/H] = --0.02 $\pm$
0.09 (random) $\pm$ 0.13 (systematic).   
This value, as well as for other element abundances, is consistent
with solar metallicity.  
If we do not take into account SO131161 and SO230062, that are not
Mayrit objects (the two no-Mayrit stars are at angular separations
$\rho >$ 30\,arcmin to the $\sigma$~Orionis centre, and might not
belong to the cluster -- Jeffries et~al. 2006; Caballero 2007, 2008b),
the new average metallicity barely varies to [Fe/H] = $-0.05 \pm 0.08
\pm 0.13$. Both metallicities would provide a distance to the cluster
of $d \sim$ 440\,pc from the results presented by Sherry et~al.~(2008). 
If we also remove from the sample the star SO420039 (that seems to be at a
longer heliocentric distance; Section~\ref{distance}) and the stars
SO211394 and SO000041 (that have relatively low lithium abundances;
Section~\ref{lithium}), then the average metallicity is [Fe/H] = $0.00
\pm 0.07 \pm 0.13$.  

\section{Summary}

We have presented WHT/WYFFOS+AF2 spectroscopy of a sample of 24 young
late-type stars of the $\sigma$~Orionis cluster, for which we have
compiled optical and near infrared photometry, radial and
rotational velocities, and effective temperatures derived with the
infrared flux method. 
Among this sample, we have selected nine late-type single stars with enough
quality spectra and effective temperatures in a suitable interval for an
accurate abundance analysis.
We have applied a $\chi^2$-minimization technique that compares a grid
a synthetic spectra with the observations. 
This method provides a determination of the stellar parameters and
metallicity. 

Including the nine stars in the selected sample, the average metallicity
for the $\sigma$~Orionis cluster is [Fe/H] $= -0.02\pm0.09\pm0.13$
(random and systematic errors), and thus in perfect agreement with
solar metallicity.
The element abundances of aluminium, silicon, calcium, and nickel also
show solar values. We have also determined the lithium
abundances of the selected nine stars.
Except for two stars (SO230062 and SO210434), the
lithium abundances are similar to or slighty lower than the
``cosmic'' $^7$Li abundance, within the error bars. 
These two stars show a lithium abundance of $\log
\epsilon(\mathrm{Li}) = 3.6-3.7$, although almost consistent with the 
``cosmic'' $^7$Li abundance within the error bars.

The stellar parameters of the remaining selected stars are consistent
with an heliocentric distance in the range  $350 < d/{\rm pc} < 500$
and an age of $3\pm1$\,Ma for the $\sigma$~Orionis cluster. 
Finally, we have determined the mean radial velocity of the cluster
at $V_r = +28 \pm 4$\,\kmso.

To sum up, the metallicity of the $\sigma$~Orionis cluster is
solar within the uncertainties of our method. Most of the investigated
stars also display cosmic lithium abundances, consistent with their
expected very young ages, although at least a couple of them would
deserve a dedicated spectroscopic follow-up at higher resolution to
confirm or reject their membership in $\sigma$~Orionis.

\begin{acknowledgements}

We are grateful to Tom Marsh for the use of the {\scshape MOLLY} analysis
package. We thank the supporter astronomer, R. Corradi, during our observations
at the WHT.
This work has made use of the VALD database and IRAF facilities. 
J.~I. ackonowledges support from the EU contract MEXT-CT-2004-014265 (CIFIST). 
Partial financial support was provided by Spanish Ministerio Educaci\'on y
Ciencia, the Universidad Complutense de Madrid, the Spanish Virtual
Observatory under grants AyA2005--05149, AyA2005--02750, AyA2005--04286 and
AyA2005--24102--E of the Programa Nacional de Astronom\'{\i}a y Astrof\'{\i}sica
and by the Comunidad Aut\'onoma de Madrid under PRICIT project S--0505/ESP--0237
(AstroCAM), MEC/Consolider--CSD2006-0070 and
CAM/PRICIT--S--0505/ESP/0361.  

\end{acknowledgements}

\end{document}